\documentclass[11pt,a4paper]{article}
\usepackage{url}
\usepackage{lineno}
\usepackage{float}
\usepackage{caption}
\usepackage{subcaption}
\usepackage{hyperref}
\usepackage{bbold}
\usepackage{multirow}
\usepackage{array}
\usepackage{balance}
\usepackage{color}
\usepackage{graphicx}
\usepackage{mathdots}
\usepackage[margin=1in]{geometry}
\usepackage[T1]{fontenc}
\usepackage[utf8]{inputenc}
\usepackage{authblk}
\usepackage{amsmath}
\usepackage{bm}
\usepackage{makeidx}
 
\newcolumntype{x}[1]{>{\centering\arraybackslash\hspace{0pt}}m{#1}}
\newcommand{\comment}[1]{}
\definecolor{Orange}{rgb}{1,0.5,0}
\definecolor{Red}{rgb}{1,0,0}

\usepackage{makeidx}  

\title{Estimating the Benefits of Electric Vehicle Smart Charging at Non-Residential Locations:\\ A Data-Driven Approach}
\author[1,2]{Emre Can Kara\thanks{eckara@lbl.gov}}
\author[1]{Jason S. Macdonald}
\author[1]{Douglas Black}
\author[2]{Mario B{\'e}rges}
\author[3]{Gabriela Hug}
\author[1]{Sila Kiliccote}
\affil[1]{Environmental Energy Technologies Division, Lawrence Berkeley National Laboratory}
\affil[2]{Civil and Environmental Engineering, Carnegie Mellon University}
\affil[3]{Electrical and Computer Engineering, Carnegie Mellon University}

\begin{document}
\maketitle

%

%

\begin{abstract} 
In this paper, we use data collected from over 2000 non-residential electric vehicle supply equipments (EVSEs) located in Northern California for the year of 2013 to estimate the potential benefits of smart electric vehicle (EV) charging. We develop a smart charging framework to identify the benefits of non-residential EV charging to the load aggregators and the distribution grid. Using this extensive dataset, we aim to improve upon past studies focusing on the benefits of smart EV charging by relaxing the assumptions made in these studies regarding: (\emph{i}) driving patterns, driver behavior and driver types; (\emph{ii}) the scalability of a limited number of simulated vehicles to represent different load aggregation points in the power system with different customer characteristics; and (\emph{iii}) the charging profile of EVs. First, we study the benefits of EV aggregations behind-the-meter, where a time-of-use pricing schema is used to understand the benefits to the owner when EV aggregations shift load from high cost periods to lower cost periods. For the year of 2013, we show a reduction of up to 24.8\% in the monthly bill is possible. Then, following a similar aggregation strategy, we show that EV aggregations decrease their contribution to the system peak load by approximately 40\% when charging is controlled within arrival and departure times. Our results also show that it could be expected to shift approximately 0.25kWh ($\sim$2.8\%) of energy per non-residential EV charging session from peak periods (12PM-6PM) to off-peak periods (after 6PM) in Northern California for the year of 2013. 
\end{abstract}

\section{Introduction}
 A recent analysis identifying the infrastructure and technology needs to meet California's greenhouse gas (GHG) reduction goals for 2050 shows that the electrification of the transportation system plays a significant role in reaching these goals. In order to achieve the 80\% reduction target in electrification, most of the direct fuel uses in buildings, transportation and industrial processes must be electrified. Among these, electrification of transportation yields the largest share of GHG reduction, where 70\% of the vehicle miles traveled should be by electrically powered vehicles~\cite{williams2012technology}. A study by the Electric Power Research Institute (EPRI)~\cite{duvall_environmental_2007} also suggests that electric vehicles will constitute a rather significant 35\% of the total vehicles in the US by 2020. \\
 
This rapid growth in the electrification of transportation presents significant challenges as well as opportunities to the operation of today's power system. When considered as inflexible loads, EVs will increase the current electricity demand significantly, intensifying the stress on the electric power system and pushing it closer to its limits. Research suggests that this is the case for low penetrations of EVs~\cite{liu_survey_2011,wu2011electric,weiller_plug_2011}. However, when considered as flexible resources, where EV charging is controlled by direct or indirect strategies, EVs promote the reliable operation of the power grid~\cite{lopes_integration_2011,galus_role_2012, guille2009conceptual}, while also providing additional revenue streams that can be used towards the electrification of transportation~\cite{liu_survey_2011, galus_role_2012,kempton2005vehicle}. This is particularly important considering the expected increase of renewable generation sources in the generation portfolio of many states in the U.S., as smart EV charging may provide the means to balance the intermittency of these resources. \\

A number of recent studies aim to understand the adaptation needs of the existing operational control mechanisms to realize smart charging, and often propose novel planning and control approaches. These approaches can be grouped into \emph{direct} and \emph{indirect} control approaches~\cite{galus_role_2012}. In direct control approaches, the control actions are realized without the vehicle owner in the control loop.  Often, load aggregations are created to increase the size of the resource so it can offer economic benefits to the aggregator~\cite{guille2009conceptual, brooks2001integration}. In~\cite{martin2012direct}, for example, the authors propose a direct load control strategy to provide vehicle-to-grid services for 3 different predefined mobility patterns. In~\cite{su2012performance}, the authors conduct a simulation study for 3000 EVs parked at a municipal parking lot and evaluate the real-time performance of a direct control approach, which maximizes the expected state of charge of the EV aggregation in the next time step subject to mobility constraints. In~\cite{he2012optimal}, the authors develop an optimal direct control scheme based on global charging costs. The authors compare the proposed direct control scheme to the local scheduler in a simulation environment including 100-400 EVs. The arrival times of the EVs, the charging periods, and the initial energies of EVs are assumed to have a uniform distribution. \\

The authors of~\cite{brooks2001integration} discuss various services that can be provided by electric vehicles, including peak shaving, regulation, voltage control, and reserves, and many studies have quantified the benefits of smart charging from various stakeholder perspectives~\cite{rotering_2011_optimal, tomic2007using,kempton2008test}. In~\cite{brooks2001integration}, the authors demonstrate a proof of concept regulation case study. In~\cite{rotering_2011_optimal}, the authors estimate that smart charging will reduce the daily electricity costs of a plug-in hybrid EV by \$0.23. They also identify daily profits for the individual driver when the charging of the vehicles can be regulated. The economic benefits of fleets that participate in specific markets have also been extensively studied. For example, in~\cite{tomic2007using}, 352 vehicles are used to estimate the economic potential of fleets when providing regulation up and down services using historical prices obtained from California Independent System Operator (ISO). In~\cite{finn2012demand}, the authors use historical market data and charging data collected from an EV located in a residential household to investigate financial savings and peak demand reduction. The authors conclude that the peak EV demand can be reduced by up to 56\%.\\

In this paper, we primarily focus on direct control approaches---in particular, centralized smart charging of EV aggregations---and we create a variety of case studies to investigate the potential benefits of smart charging to different stakeholders. To develop these case studies, we use data collected from over 2000 non-residential electric vehicle supply equipments (EVSEs) located throughout 190 zip code regions in Northern California spanning one year. To the best of our knowledge, this is the first study that uses such an extensive dataset on EV charging. First, we analyze over 580,000 charging sessions to investigate the trends in load flexibility and infrastructure use in the dataset. Next, we create virtual aggregation points (VAP) in which a combination of the EVSEs is assumed to be fed by the same distribution feeder. The VAPs mostly coincide with Pacific Gas and Electric Company's (PG\&E) sub-load aggregation points (sub-LAPs). Additional details regarding this relationship is provided in Section~\ref{evsec:dataset}. We introduce a smart charging framework to estimate the benefits of smart EV charging to various stakeholders in each VAP. As an initial case study, we investigate the potential benefits of EV aggregations operated under a single owner, where a time-of-use pricing scheme is used to estimate economic benefits to the owner via shifting load from high cost periods to lower cost periods. Then, we create a case study where EV aggregations are used to decrease their current contribution to the system-level peak load.\\

The motivation for this study is threefold: (\emph{i}) Most of the work investigating the potential of smart charging of EVs is based on assumptions made regarding trip and customer characteristics. For example, in~\cite{sundstrom_flexible_2012}, the authors use a fleet which includes commuter cars, family cars and taxis with predetermined departure and arrival locations randomly selected from a limited number of alternatives. In~\cite{dietz_economic_2011}, the authors use data from driving surveys that reflect the driving behavior of people using internal combustion engine cars. They assume that the driving behavior of an EV owner will be similar to that of an internal combustion engine car owner. The dataset used in this study allows us to extract trip and customer characteristics, hence no such assumptions are needed on these characteristics. (\emph{ii}) Often, a limited number of vehicles and mobility patterns are used in fleet-based studies to capture the most likely driving scenarios. For example, in~\cite{martin2012direct}, the authors develop a proof of concept strategy and show cost benefits for 50 EVs with 3 different pre-defined mobility patterns. Although the exact number of EVs are not available in the dataset used in this study, the number of charging sessions (over 580,000) and the fact that these charging sessions are spread throughout the year ensure that a representative population of non-residential charging is studied. (\emph{iii}) The individual charging profile of an EV is often represented by a typical constant-voltage, constant-current curve for certain battery chemistries, or more simply by a constant charging power~\cite{galus_role_2012}. For example, in~\cite{deilami_realtime_2011}, the charging power is assumed to be fixed at 4.4kW, whereas in~\cite{shaaban_real_2014}, the authors use the charging profile of a typical lithium-ion battery pack obtained from~\cite{marra_demand_2012}. The dataset used in this study includes time series of power measurements obtained every 15 minutes for each charging session. Hence, no assumptions are made on charging profiles of the vehicles, and individual charging data is available for each charging session.\\

The remainder of the paper is organized as follows: Section~\ref{evsec:dataset} introduces the dataset and discusses the load flexibility and infrastructure use trends obtained from the dataset. Section~\ref{evsec:methodology} presents the smart charging strategy used in this study. Specifically, it discusses the framework and the underlying assumptions made when estimating the benefits to different stakeholders. Sections~\ref{evsec:chargingowner} and~\ref{evsec:dso} describe the case studies completed in this research. Finally, Section~\ref{evsec:conc} discusses the conclusions and opportunities for future work.     

\section{Dataset}
\label{evsec:dataset}
\begin{figure}[h!]
\centering
\includegraphics[width=0.9\textwidth]{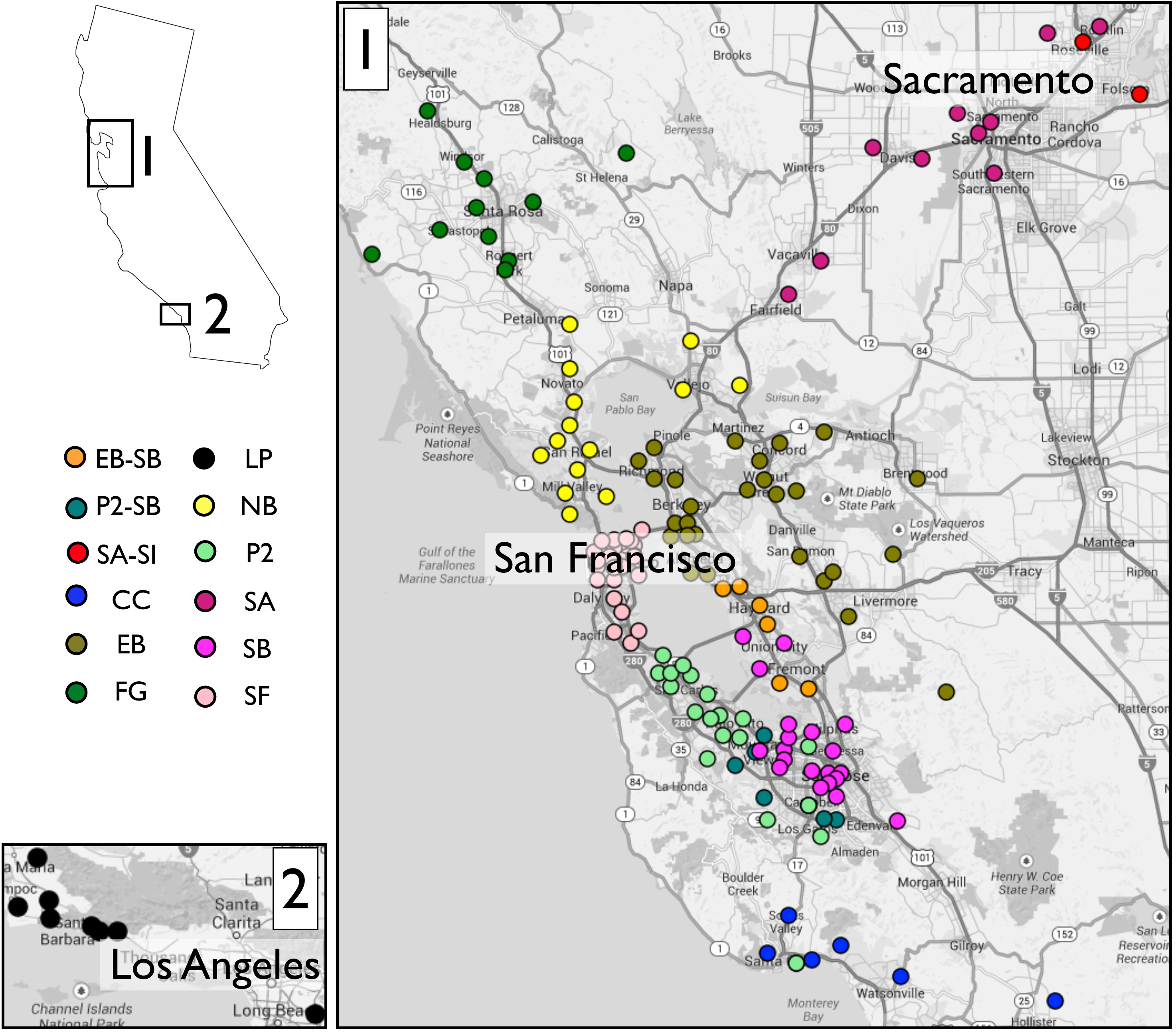}
\caption{Centroids of zip code regions forming the VAPs}\label{fig:location}
\end{figure}

The data used in this study is collected from individual EVSEs located in 16 different sub-LAPs in PG\&E's territory for the year of 2013. For each charging session (i.e. from plug-in to departure of an EV), the EVSEs report the start and end period of the charging, the plug-in and departure time stamps, the average power, and the maximum power (measured every 15 minutes), as well as the charging port type, the location (zip-code level), and the non-residential building category. Since the dataset includes the location information based on zip codes and some zip codes are fed by multiple sub-LAPs, we create virtual aggregation points (VAPs) for the zip codes that are fed by multiple sub-LAPs. This is done by combining the sub-LAPs' identifiers. Table~\ref{tab:regions} presents the final list of VAPs in the dataset and total number of zip code regions forming each of these VAPs, the total number of charging sessions, and the average number of daily charging sessions in each VAP. Figure~\ref{fig:location} depicts the centroids of the zip code regions forming the considered VAPs.\\

\begin{figure}
\centering
\includegraphics[width=0.9\textwidth]{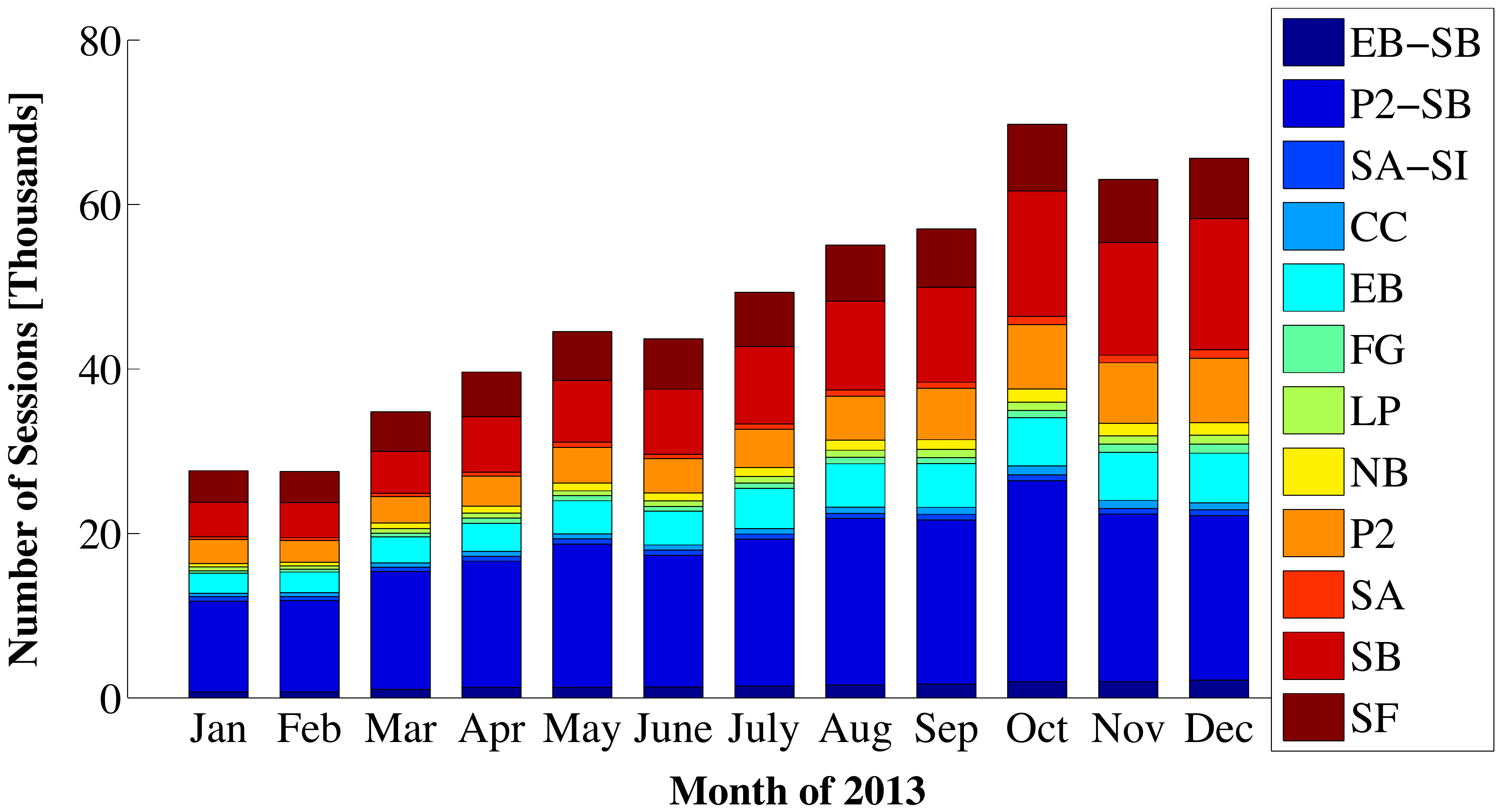}
\caption{Number of sessions per month}\label{fig:NumSessionsPerMonth}
\end{figure}

In this study, we use data from VAPs with an average of 20 or more charging sessions per day. These VAPs are indicated in \textbf{bold} in Table~\ref{tab:regions}. Figure~\ref{fig:NumSessionsPerMonth} also shows the total number of charging sessions per month for each VAP used in this study. Over the course of 2013, the total number of charging sessions approximately doubles.\\

\begin{figure}
\centering
\includegraphics[width=0.9\textwidth]{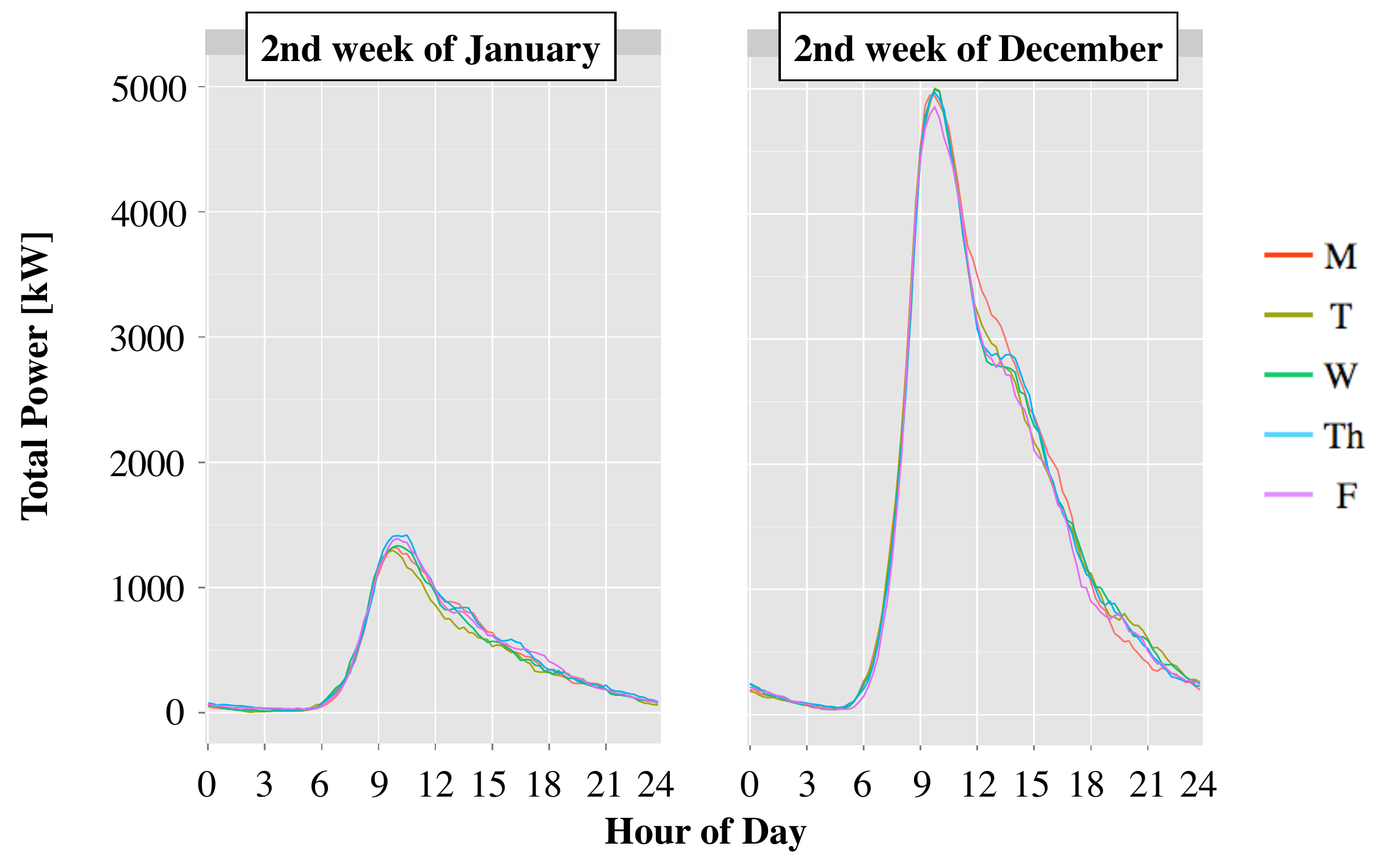}
\caption{Load shapes for January and December for all the VAPs}\label{fig:sampleLoadShapes}
\end{figure}

Figure~\ref{fig:sampleLoadShapes} shows the combined load profiles of VAPs for the second weeks of January and December. The impact of the growth in charging session is reflected on the daily load profile of the loads. Moreover, the peak non-residential EV load occurs between 9AM and 11AM, and it more than triples from January to December of 2013.   
\begin{table}
\centering
\begin{tabular}{x{0.7in}|x{1.8in}|x{1in}|x{1in}|x{1in}} 
\bf{VAP} & \bf{Region} & \bf{ \# of zip code regions} & \bf{\# of charging sessions} & \bf{\# of charging sessions per day}\\
\hline
\hline
\bf{P2-SB} & Peninsula \& South Bay & 7 & 207501 & \bf{568.50}\\
\bf{SB} & South Bay  & 21 & 112250 & \bf{307.53}\\
\bf{SF} & San Francisco & 30 & 72996 & \bf{199.99}\\
\bf{P2} & Peninsula &17 & 59252 & \bf{162.33}\\
\bf{EB} & East Bay & 27 & 52700 & \bf{144.38} \\
\bf{EB-SB} & East Bay \& South Bay & 6 & 16902 & \bf{46.31}\\
\bf{NB} & North Bay &14 & 12346 & \bf{33.82}\\
\bf{LP} & Los Padres &8 & 9035 & \bf{24.75} \\
\bf{CC} & Central Coast &15 & 8428 & \bf{23.09} \\
\bf{SA} & Sacramento Valley & 11 & 7787 & \bf{21.33}\\
\bf{FG} & Geysers & 11 & 7918 & \bf{21.69} \\
\bf{SA-SI} & Sacramento V. \& Sierra & 2 & 7465 & \bf{20.45}\\
\hline
CC-P2 & Central Coast \& Peninsula & 2 & 6778 & 18.57\\
FG-NB & Geysers \& North Bay & 4 & 3845 & 10.53\\
F1 & Fresno &4 &  377 & 1.03 \\
NV & North Valley &1 & 336 & 0.92\\
ST & Stockton & 3 & 244 & 0.67\\
FG-NC & Geysers \& North Coast & 1 & 246 & 0.67\\
SI & Sierra & 2 & 181 & 0.50\\
SN & San Joaquin & 1 & 134 & 0.37\\
HB & Humboldt & 1 & 101 & 0.28 \\
P2-SF & Peninsula \& San Francisco&1 & 73 & 0.20\\
NC & North Coast &1 & 15 & 0.04\\
\end{tabular}
\caption {VAPs used in this study} \label{tab:regions}
\end{table}

\subsection{Load Flexibility and Infrastructure Use}
\begin{figure}
\centering
\includegraphics[width=0.9\textwidth]{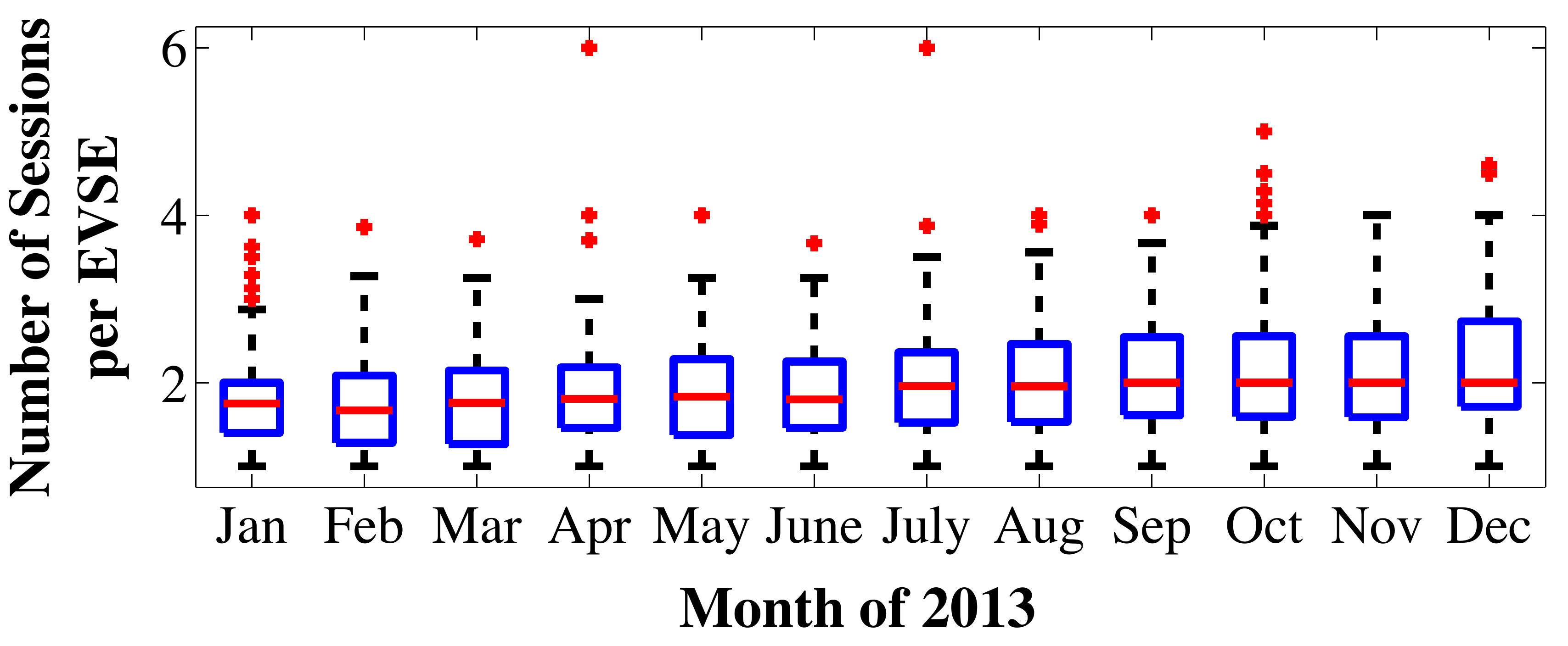}
\caption{Average number of sessions per unique EVSE per day}\label{fig:events_per_evse}
\end{figure}

To gain further insight into the dataset and to understand the distribution of charging sessions and the use of EVSEs in different regions, we analyze the charging sessions obtained from the VAPs marked in bold Table~\ref{tab:regions}. The infrastructure use, $I_{use}$, in each VAP is represented by the average number of charging sessions $N_{sessions}$ per EVSE and calculated for every business day of 2013. Formally:

\begin{equation}
I_{use}=\frac{N_{sessions}}{N_{EVSE}}
\end{equation}

\noindent where $N_{EVSE}$ is the number of EVSEs.
Figure~\ref{fig:events_per_evse} depicts the box plots of the infrastructure use within 2013 for all of the VAPs. For each month of 2013, a box plot is created to represent the distribution of the $I_{use}$ values calculated for every business day of the month. The median value of infrastructure use is marked with a red line in each box plot, and the boundaries of the box depict the 25th and 75th percentiles. The whiskers correspond to the 99th percentiles assuming the distributions per each month are normal. The median infrastructure use increases in all VAPs from 1.8  to 2.1 sessions per EVSE from January to December. We believe that this is due to the fact that the demand has increased faster than the number of EVSEs.  \\

In addition to the infrastructure use, we investigate the load flexibility in each VAP. The load flexibility depends on the charging duration $d_{\text{charge}}$ and the overall duration of each charging session $d_{\text{session}}$. Formally, we define the load flexibility $l_{\text{flex}}$ as the ratio of the duration that a car is plugged but not charging to the overall session duration:

\begin{equation}
l_{\text{flex}}=\frac{d_{\text{session}}-d_{\text{charge}}}{d_{\text{session}}}
\end{equation}

\begin{figure}
\centering
\includegraphics[width=0.9\textwidth]{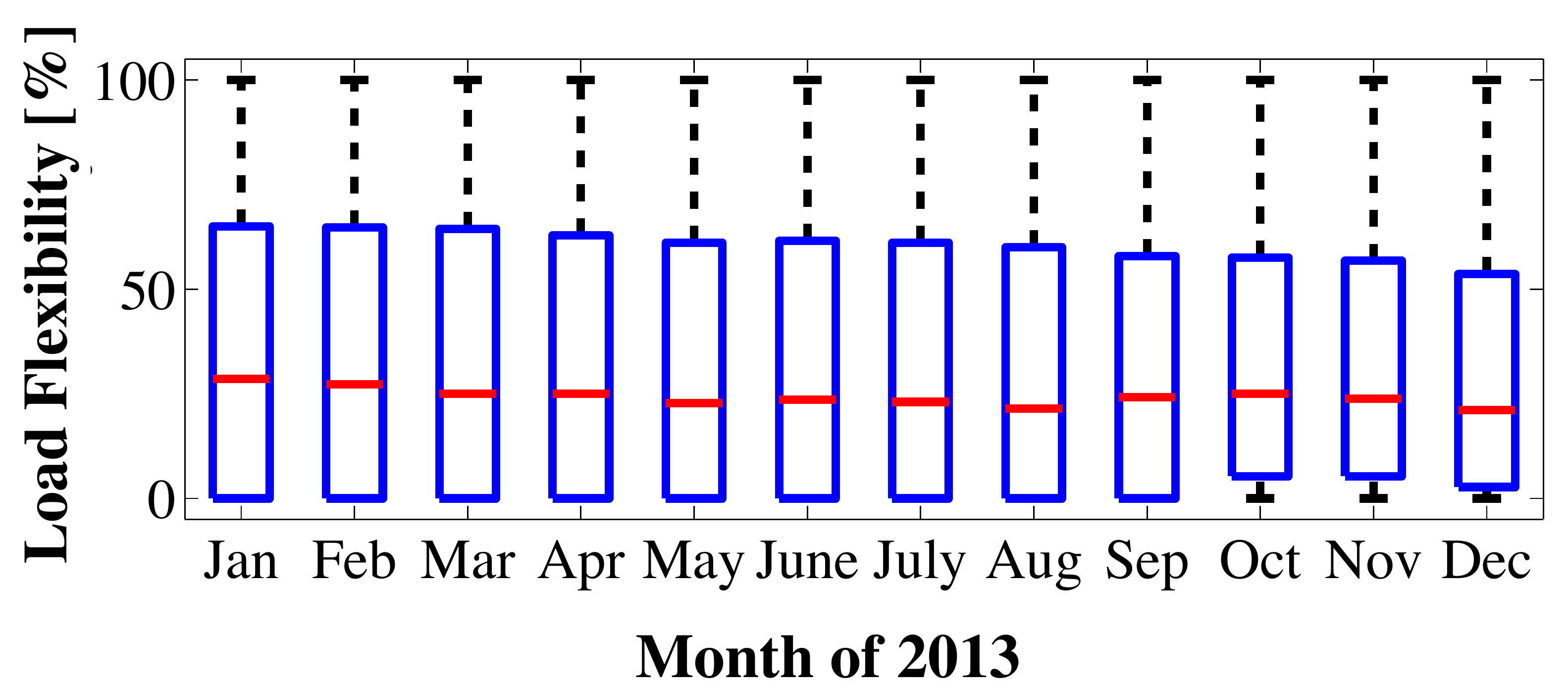}
\caption{The variation in load flexibility}\label{fig:loadflex}
\end{figure}

Figure~\ref{fig:loadflex} depicts the load flexibility for all VAPs by month. As Figure~\ref{fig:loadflex} suggests, the load flexibility decreases slowly as the number of charging sessions per EVSE increases. Also, most of the distributions have a slight positive skew. The size of the box representing the 25th and 75th percentiles is also decreasing with time, suggesting an increase in skewness.\\

\begin{figure}
        \centering
        \begin{subfigure}[b]{0.5\textwidth}
                \includegraphics[width=\textwidth]{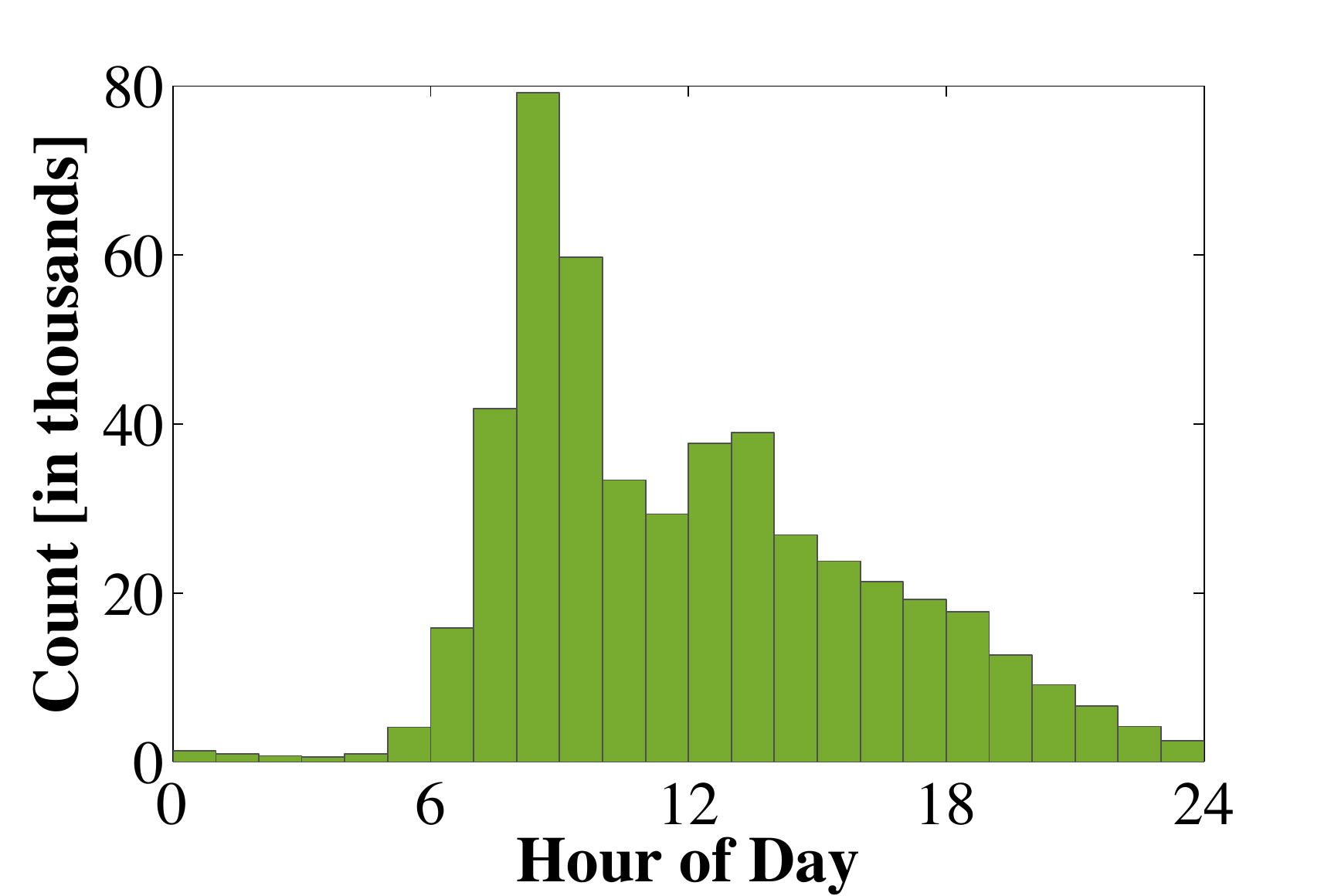}
                \caption{Distribution of arrival time}
                \label{fig:startHour}
        \end{subfigure}%
        ~ 
        \begin{subfigure}[b]{0.5\textwidth}
                \includegraphics[width=\textwidth]{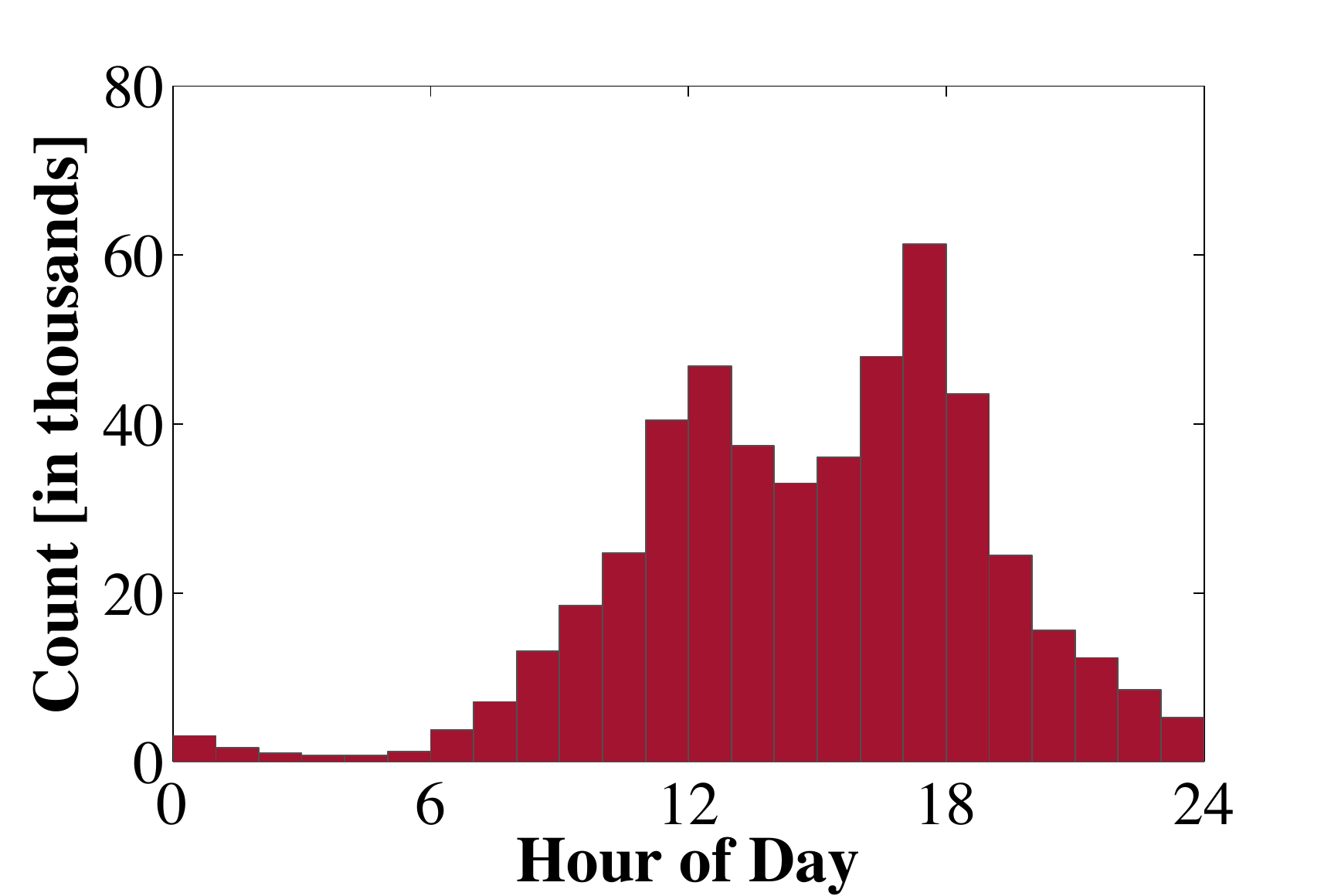}
                \caption{Distribution of departure time}
                \label{fig:endHour}
        \end{subfigure}
        \caption{Arrival and departure time characteristics}
\end{figure}

The load flexibility metric shows the charging duration relative to the session duration; however, it does not capture when the charging sessions occur. The start and end times of the charging sessions play a key role when estimating the benefits of EV aggregations to the power system. To put these two variables into perspective, we show a histogram of arrival (i.e. session start) and departure (i.e. session end) times in Figures~\ref{fig:startHour} and~\ref{fig:endHour}, respectively.\\

As can be seen in Figures~\ref{fig:startHour} and~\ref{fig:endHour}, most of the charging sessions start within the 7AM-10AM period and often end within the 5PM-7PM period. Considering these loads are currently uncontrolled (i.e.~they immediately start charging when they are plugged in), they coincide with the typical working hours of a non-residential location. These figures suggest that employees or customers arrive in the morning and plug in their vehicles. Some leave around noon and come back, and most leave work between 4PM and 7PM. 

\section{Smart Charging Strategy}
\label{evsec:methodology}
In this section, we introduce the proposed smart charging methodology. In particular, we describe the general optimization strategy used to obtain the charging schedules for each charging session.\\

The goal of the proposed smart charging framework is to reschedule the power time series measured in discrete time slots $[1,\hdots,K]$ for any charging session in a population of EVs, $[P_{1},P_{2},\hdots,P_{K}]$ such that an objective function is optimized. The objective function should capture the desired benefits from a stakeholder's perspective. While rescheduling the charging, we would like to ensure that the order of the measurements in this time series is preserved. This is because the power that EVSEs draw is dependent on the state of charge (SOC) of the EV that is being charged, and keeping the order of the measurements accounts for this dependency. In addition, we assume that the charging is preemptive; that is, the charging tasks are interruptible without any decrease in the SOC of the EV. \\

In a typical charging session, an EV starts charging when it is plugged in, and often the charging is complete before the vehicle departs. The smart charging framework proposed in this study is designed to move some of the charging to the slack time slots (i.e. the time slots where the vehicle is plugged in but the charging is completed).\\ 

For the purposes of this paper, we discretize a day into 15-minute intervals. We define the time period for the optimization within a day as the time between the start time slot $t_{start}$ and the end time slot $t_{end}$. In this period, each charging session $i$ has an arrival time slot denoted by $t^{(i)}_{a}$ and a departure time slot $t^{(i)}_{d}$. For each charging session, a column vector including the charging power time series can be created using the power measurements for every time slot in [$t^{(i)}_{a}$, $t^{(i)}_{d}$]. If necessary, the time series is zero-padded to match the size of the optimization time period [$t_{start}$, $t_{end}$]. Hence, for each EV $i$, the power time series is given as follows:

\begin{equation}
\bm{P}^{(i)}=[P^{(i)}_{1},P^{(i)}_{2},\hdots,P^{(i)}_{K}]^T
\end{equation} 

\noindent where $K$ is the total number of time slots in [$t_{start}$, $t_{end}$]. Next, for each charging session $i$, we identify $\bm{Q}^{(i)}$ whose elements $Q^{(i)}_j$ correspond to the $j^{th}$ non-zero element of $\bm{P}^{(i)}$. The goal is to reschedule the time slots $t^{(i)}_{j}$ in [$t^{(i)}_{a}$, $t^{(i)}_{d}$] corresponding to $Q^{(i)}_j$ without changing their order. We define $M^{(i)}$ as the total number of non-zero power measurements in charging session $i$ (i.e. total number of elements in $\bm{Q}^{(i)}$).\\ 

To capture the precedence and the session duration constraints we proposed above, the following formal constraints are introduced: 

\begin{equation}
\label{eq:constraints}
\left.
\begin{aligned}
& t^{(i)}_{j}\geq t_{start}\\
& t^{(i)}_{j} \leq t_{end}\\
& t^{(i)}_{j} \geq t^{(i)}_{a}\\
& t^{(i)}_{j} \leq t^{(i)}_{d}\\
& t^{(i)}_{j} < t^{(i)}_{j+1}\\
\end{aligned}
\right\} 
\begin{aligned} &\forall i \in [1,N],\\ &\forall j \in [1,M^{(i)}] \end{aligned} 
\end{equation}

\noindent The proposed constraints are constructed using a binary decision matrix to represent charging or non-charging time slots within the optimization duration. In particular, for each element $Q^{(i)}_{j}$ in $\bm{Q}^{(i)}$, we create a binary vector $x^{(i,j)}$ that includes $K$ binary decision variables. Each element in this vector represents a candidate time slot at which $Q^{(i)}_j$ could be positioned. Hence, we define row vectors $x^{(i,j)}$ $\forall i \in[1,N]$ and $\forall  j \in [1,M^{(i)}]$. The elements in these vectors are $x^{(i,j)}_{k} \in \{0,1\}$ that are defined $\forall k \in[1,K]$.\\

From these binary vectors $x^{(i,j)}$, we form a binary decision matrix  $\bm{X}^{(i)}$ for each charging session $i$ $\in$ $[1, N]$. In particular, the individual decision variables $x^{(i,j)}_{k}$ form the elements of the binary decision matrix $\bm{X^{(i)}}$ as follows:

\begin{equation}
\bm{X^{(i)}}=\begin{bmatrix} x^{(i,1)}_{1} & \hdots & x^{(i,1)}_{K} \\
 \vdots& \ddots & \vdots \\
 x^{(i,M^{(i)})}_{1} & \hdots & x^{(i,M^{(i)})}_{K} 
 \end{bmatrix}
 \end{equation}
 
\noindent Finally, we write the variables in the constraints given in~(\ref{eq:constraints}) using the binary decision variable as follows:

\begin{equation}
\label{eq:optim}
\begin{aligned}
& t^{(i)}=\bm{X^{(i)}} O,\ \text{where} \ O=\begin{bmatrix} 1\\ 2\\ \vdots\\ K\end{bmatrix} \end{aligned} 
\end{equation}

\noindent The aggregate power vector for the VAP $AP^{(d)}=\sum_{i=0}^N (\bm{P}^{(i)})$ for the day $d$ is given as follows:

\begin{equation}
AP^{(d)}=\begin{bmatrix} \bm{Q}^{(1)}\\ \bm{Q}^{(2)}\\ \vdots \\ \bm{Q}^{(N)} \end{bmatrix}^T \begin{bmatrix} \bm{X}^{(1)}\\ \bm{X}^{(2)}\\ \vdots \\ \bm{X}^{(N)}\end{bmatrix}\\
\end{equation}

For each case study proposed in this paper, we build on the general optimization framework described above, identify the objective functions to capture the benefits from each stakeholder's perspective and introduce additional constraints when necessary.
We use the Gurobi optimizer~\cite{optimization2014inc} to solve the optimization problems formulated for each case study. Due to the size of the optimization problem for certain VAPs and the number of times the optimization problem is solved to obtain values to estimate benefits for the year of 2013, a proved optimal solution is expected to be hard to reach within a reasonable time frame. For these reasons, we alter the optimality criteria by controlling the relative gap between a feasible integer solution and the general optimal solution. We set this optimality criteria to 5\% and allow early termination once a feasible solution is found.  

\section{Charging Infrastructure Owner's Perspective}
\label{evsec:chargingowner}
\indent In the first case study, our goal is to capture and maximize the benefits of smart charging from an EV charging service provider's perspective. Currently, each charging meter is independently owned by the building owner, and the consumption is billed to the building owner as part of the building's monthly bill. However, in our work, we focus only on the load resulting from EV charging, i.e.~decoupled from other loads, but aggregated over VAPs formed based on sub-LAPs. This corresponds to the situation in which the charging stations within each VAP are combined and operated under a single owner or an aggregator and the owner is charged according to a time of use (TOU) tariff structure, where shifting load from high cost periods to lower cost periods can offer some benefits to the owner. Although the current VAPs are created based on sub-LAPs, the current scale of the charging infrastructure and the number of charging sessions can easily represent a large parking structure or a campus in the future, where the EV aggregation is behind a single meter and the non-EV load is relatively steady.\\

\subsection{Problem Formulation}
In a typical TOU rate structure, there are two separate charges forming the monthly bill: the \emph{energy charges} and the \emph{demand charges}. The energy charges are calculated based on the amount of energy consumed over given time periods of the day using the corresponding hourly TOU energy rate. The demand charges are calculated based on the maximum power demand for specific time periods of the day over the course of the billing period. At the end of each billing period, the maximum demand values for the specified periods are multiplied by the demand charge rates and added to the overall energy charge.\\ 

In order to model a similar rate structure in the proposed smart charging framework, we define $EC^{(d)}$ as the energy charge for day $d$ of a month with $D$ days (i.e. $d \in [1,\hdots, D]$). Then, we define $DC_h$ as the demand charges for each time period $h$ of the day of any month. For example, in PG\&E's E-19 TOU rate structure, for winter billing periods, the demand charges are calculated based on 2 time periods \emph{part-peak} (i.e.~8:30AM-12:00PM \& 6:00PM-09:30PM) and \emph{off-peak} (i.e.~09:30PM-08:30AM)~\cite{e19}. Formally, the monthly bill for the owner is therefore given by:

\begin{equation}
\label{eq:cost}
f (DC_h,EC^{(d)})=\sum_{\forall h} DC_h +\sum_{\forall d}EC^{(d)}
\end{equation}

The energy charges $EC^{(d)}$ can easily be incorporated into the  proposed daily optimization routine. Defining $ER$ as a column vector reflecting the price of energy for each time slot $j$, $EC^{(d)}$ for any day $d$ in a billing period is given by:

\begin{equation}
EC^{(d)}=AP^{(d)}ER\\
\end{equation}

\noindent For time period $h$ within day $d$, a subset of the entire daily aggregate power vector $AP^{(d)}$ is needed and is referred to as  $AP^{(d)}_h$.\\

In order to minimize the cost function given in~(\ref{eq:cost}), the maximum demand for the daily time periods $h$ must be accurately known beforehand for the entire month. However, in a real life scenario, this is not a valid assumption. To incorporate demand charges into the proposed daily smart charging framework, we therefore propose the following strategy for the owner: for each day $d$, we define the peak aggregate power values for each period $h$ as $AP_{peak,h}^{(d)}$. Since the historic $AP_{peak,h}$ values for each day in $[1,\hdots,d-1]$ are available to the main scheduler, we can define the maximum of the historic $AP_{peak,h}$ values until $d-1$ as follows:
 
\begin{equation}
AP_{max,h}^{(d-1)}=max(AP_{peak,h}^{(1)},\hdots,AP_{peak,h}^{(d-1)})
\end{equation}

Using the above definition, the monthly demand charges can be calculated at the end of the month based on $AP_{max,h}^{(D)}$ and the demand rates $DR_{h}$ for each period as:

\begin{equation}
DC_{h}=AP_{max,h}^{(D)} DR_{h}
\end{equation}

As we move from one day to the next, we try to limit the demand charges based on the maximum daily demands occurred up to the current day. At the beginning of the billing period, we start with no knowledge of the historical peak values, and we keep track of the maximum historical value up to day $d$. This strategy can be represented by incorporating the maximum value of the peak values $AP_{max,h}^{(d)}$ for time period $h$ and day $d$ as decision variables into the following optimization problem:

\begin{equation*}
\begin{aligned}
& \underset{\bm{X}^{(i)},AP_{max,h}^{(d)}}{\text{minimize}}
&AP_{max,h}^{(d)} DR_h+EC^{(d)}\\
\end{aligned}
\end{equation*}

\noindent subject to~(\ref{eq:constraints}) and the following additional constraints:

\begin{equation}
\label{eq:DCoptim}
\left.
\begin{aligned}
& AP_{max,h}^{(d-1)} \le AP_{max,h}^{(d)}\\
& AP_{h}^{(d)} \le AP_{max,h}^{(d)} \\ \end{aligned} 
\right\} \begin{aligned} &\forall h \in [1,TP] \end{aligned} 
\end{equation}

Note that with~(\ref{eq:DCoptim}), we ensure that the current maximum $AP_{max,h}^{(d)}$ is more than or equal to the maximum historical value $AP_{max,h}^{(d-1)}$ for period $h$. By definition, this allows for  the tracking of the maximum value up to that day. In addition, these maximum values set the day based on which the demand charges will be calculated. If none of the current peak values exceeds the historical maximum values, the demand charges for each period $h$ are not set by the current day $d$. \\
\subsection{Case Study}
For the purposes of this paper, we use the demand and energy rates from PG\$E's E-19 TOU rate structure~\cite{e19}. The E-19 rate structure gives the owner the option to manage their electric costs by shifting load from high cost periods to lower cost periods. Detailed information on E-19 is given in Table~\ref{tab:rates}. The summer period starts with May 1st and ends October 31st, and the winter period includes the remaining months of the year. This rate is for non-residential customers in PG\&E's territory with highest demand exceeding 499 kW for three consecutive months. \\
\begin{table}
\centering
\begin{tabular}{p{2.4in} p{0.7in} x{1.3in}} 
\bf{Demand Charges} & \bf{\$/kW} & \bf{Time Period} \\
\hline
Max. Peak Demand Summer &\$19.71253&12:00PM-6:00PM\\
Max. Part-Peak Demand Summer & \$4.07  & 8:30AM-12:00PM \& 6:00PM-09:30PM\\
Max. Demand Summer & \$12.56 & Any time\\
Max. Part-Peak Demand Winter & \$0.21& 8:30AM-09:30PM \\
Max. Demand Winter & \$12.56 & Any time\\
&&\\
\bf{Energy Charges} &\bf{\$/kWh} & \bf{Time Period}\\
\hline
Peak Summer & \$0.16253 & 12:00PM-6:00PM\\
Part-Peak Summer & \$0.11156  & 8:30AM-12:00PM \& 6:00PM-09:30PM\\
Off-Peak Summer & \$0.07818 & 09:30PM-08:30AM\\
Part-Peak Winter & \$0.10479 & 08:30AM-09:30PM \\
Off-Peak Winter & \$0.08200 & 09:30PM-08:30AM\\
\end{tabular}
\caption {E-19 rate structure~\cite{e19}} \label{tab:rates}
\end{table}

To evaluate the benefits of smart charging when the EV aggregation has a single bill calculated on a TOU tariff, we first calculate the current bill under this tariff but without smart charging. Then, we use the proposed optimization strategy to schedule the loads in a way that minimizes the customer's monthly bills, and we report each monthly bill calculated for each VAP and the contributions from energy and demand charges in the bill. 

\subsection{Results}
\begin{figure}
\centering
		\includegraphics[width=0.9\textwidth]{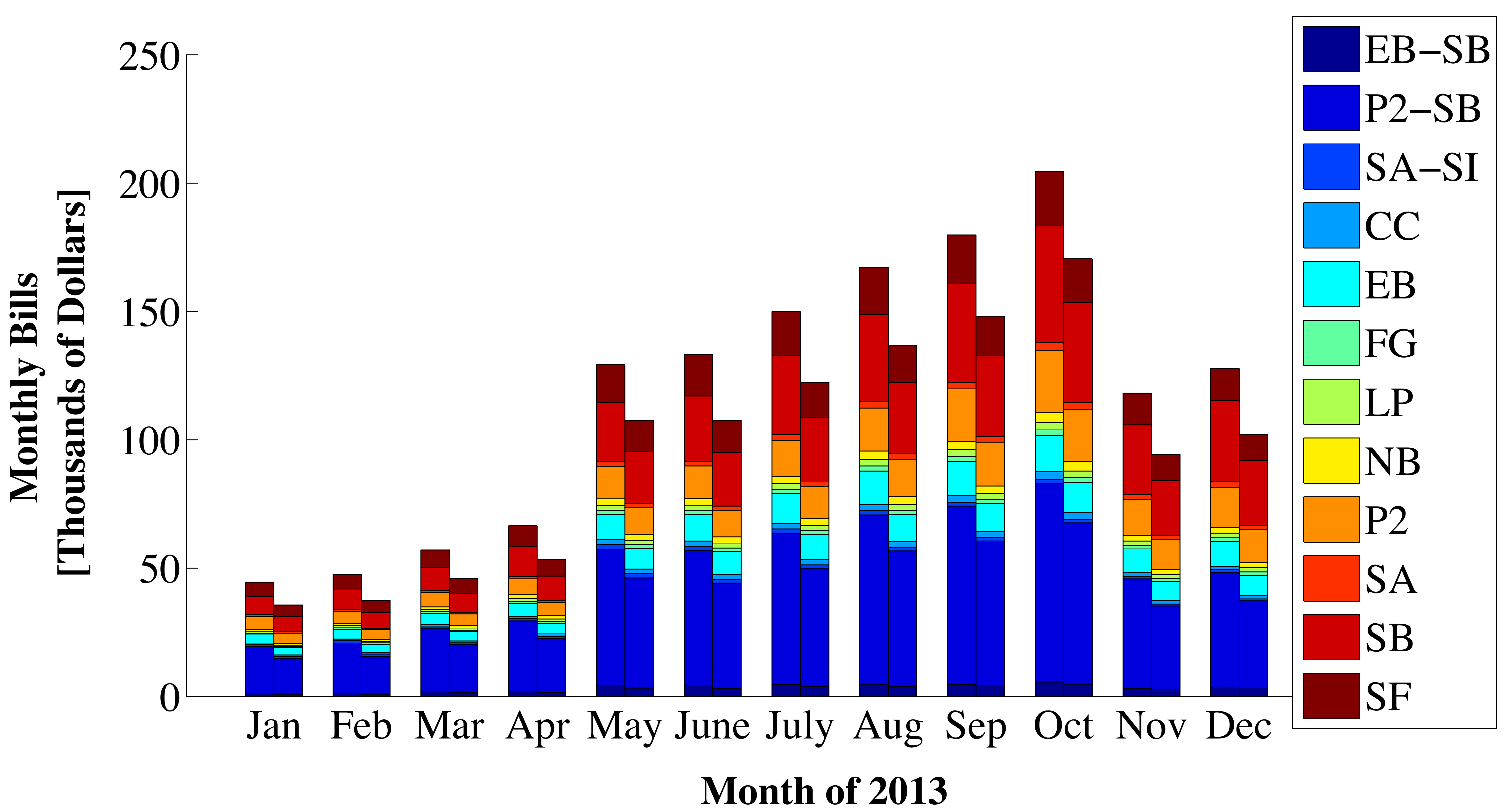}
		\caption{Monthly bills calculated with E-19. The left bar for each month shows the current bill, and the right bar shows the optimized bill.}\label{fig:MonthlyBills}
\end{figure}

Figure~\ref{fig:MonthlyBills} shows the sum of monthly bills calculated in dollars for all of the VAPs. For each month, the left bar shows the current bill, and the right bar shows the optimized bill for the month. It is obvious that the difference between the summer and winter rates impacts the aggregate monthly bill. The increase within the winter and the summer period is due to the increase in the number of charging sessions over the year. \\
\begin{figure}
\centering        
	\begin{subfigure}[a]{1\textwidth}
		\includegraphics[width=0.9\textwidth]{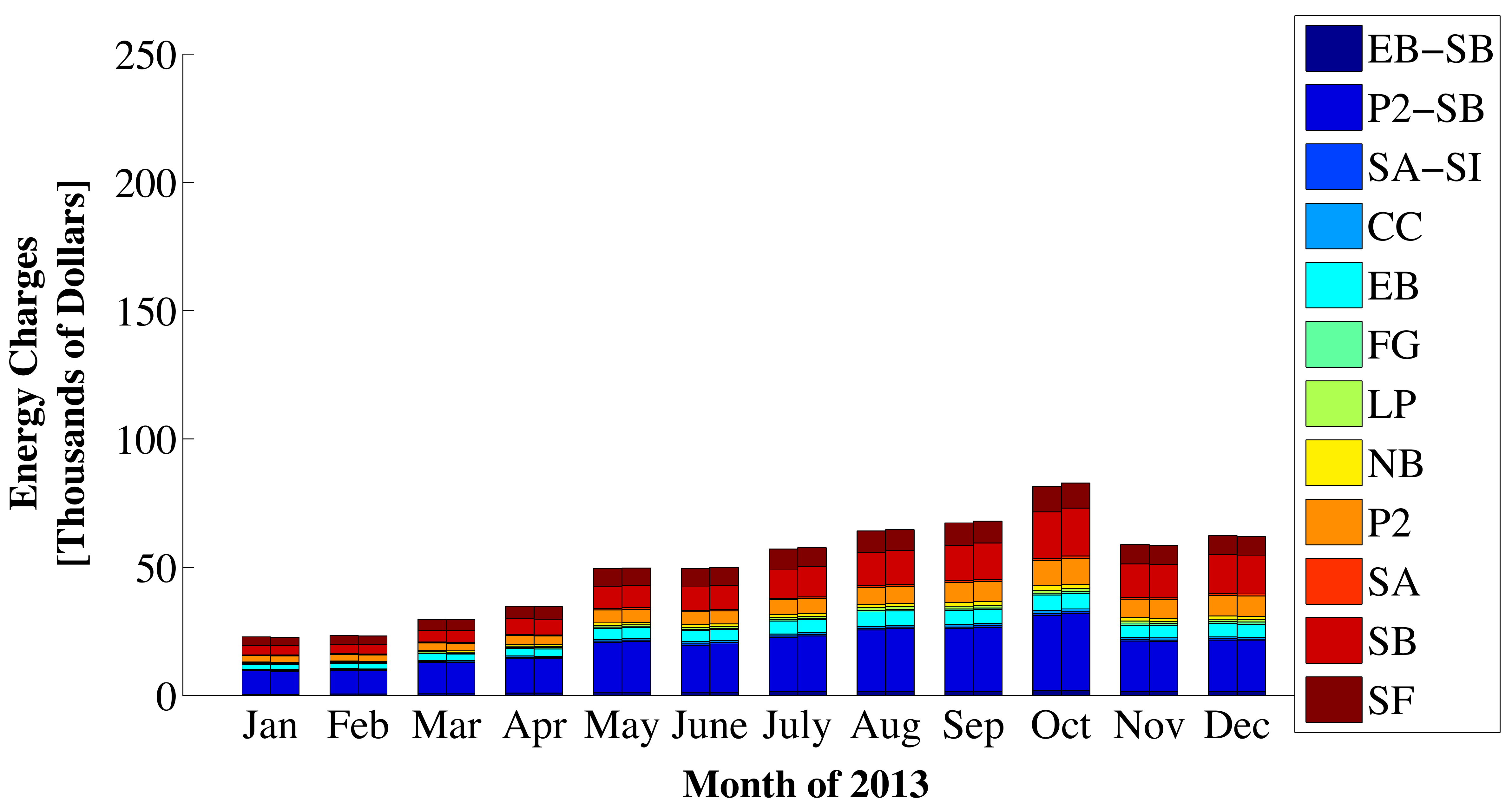}
		\caption{Monthly energy charges calculated with E-19}\label{fig:MonthlyEnergyCharges}
         \end{subfigure}
         \begin{subfigure}[b]{1\textwidth}
		\includegraphics[width=0.9\textwidth]{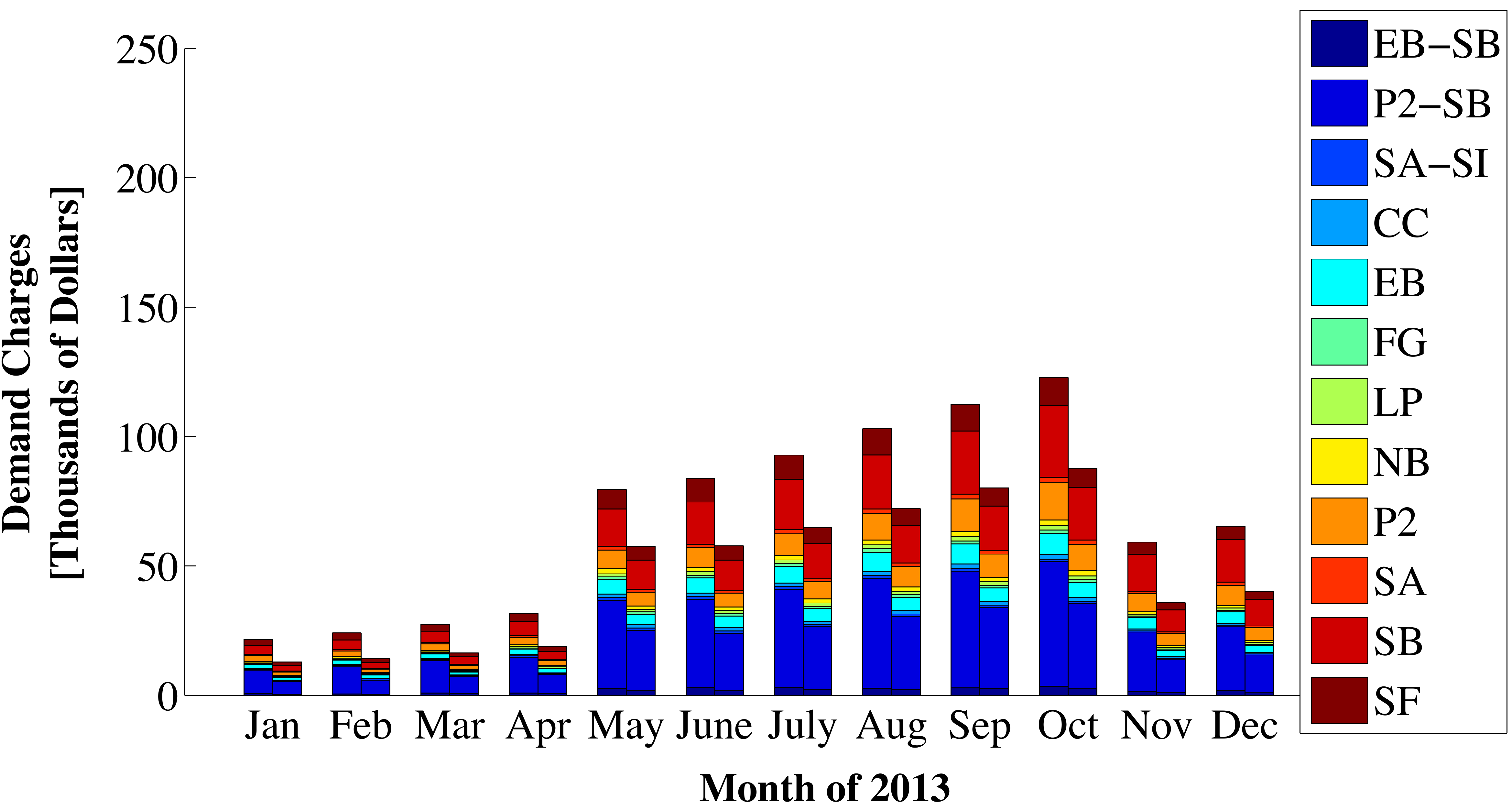}
		\caption{Monthly demand charges calculated with E-19}\label{fig:MonthlyDemandCharges}
         \end{subfigure}%
        ~ 
\caption{Decomposition of monthly bills to energy and demand charges. In each figure, the left bar shows the current charges, and the right bar shows the optimized charges for each month.}
\end{figure}

Figures~\ref{fig:MonthlyEnergyCharges} and~\ref{fig:MonthlyDemandCharges} show the total energy and demand charges, respectively, over all LAPs. The cumulative energy charges increase slightly for the summer months when using smart charging, whereas there is a significant drop in the demand charges. This suggests that the peak load of the EVs is shifted from the morning partial-peak period to the peak-period. This shift is still beneficial because the increase in the energy charges is significantly lower than the decrease in the demand charges.\\     

\begin{table}
\centering
\small
\begin{tabular}{|x{1cm}|x{1.3cm}|x{1.2cm}x{1.49cm}|x{1.7cm}|x{1.2cm}x{1.1cm}x{1.1cm}|} 
\hline
 \multirow{2}{*}{\bf{VAP}} & \multirow{2}{*}{\bf{Period}} & \multicolumn{2}{c|}{\bf{Bill [dollars]}} &{\bf{Reduction}}&\multicolumn{3}{c|}{\bf{Reduction [\%]}} \\
  \cline{3-4}\cline{6-8}
&&Current&Optimized&\bf{[dollars /session]}& DC & EC&Total\\
\hline
\multirow{2}{*}{P2-SB}&Summer&63001&50395&0.65&20.86\%&-0.85\%&20.01\%\\
&Winter&29603	&22575&0.46&23.41\%&0.33\%&23.74\%\\
\cline{2-8}
\multirow{2}{*}{EB-SB}&Summer&4588&3788&0.52&16.96\%&0.49\%&17.45\%\\
&Winter&2092&1724&0.28&17.23\%&0.36\%	&17.59\%\\
\cline{2-8}
\multirow{2}{*}{SA-SI}&Summer&1645&1413&0.36&13.80\%&0.30\%&14.10\%\\
&Winter&828&752&0.13&9.06\%&0.12\%&9.18\%\\
\cline{2-8}
\multirow{2}{*}{CC}&Summer&2365&2178&0.24&7.34\%&0.57\%&7.91\%\\
&Winter&1037&896&0.22&13.31\%&0.29\%&13.60\%\\
\cline{2-8}
\multirow{2}{*}{EB}&Summer&12033&10003&0.41&16.44\%&0.43\%&16.87\%\\
&Winter&5874&4868	&0.26&16.66\%&0.47\%&17.13\%\\
\cline{2-8}
\multirow{2}{*}{FG}&Summer&1803&1568&0.33&11.98\%&1.05\%&13.03\%\\
&Winter&920&807&0.18&11.82\%&0.46\%&12.28\%\\
\cline{2-8}
\multirow{2}{*}{LP}&Summer&2370&2135&0.29&9.37\%&0.55\%&9.92\%\\
&Winter&1141&1002	&0.20&11.88\%&0.30\%&12.18\%\\
\cline{2-8}
\multirow{2}{*}{NB}&Summer&3136&2865&0.23&8.16\%&0.49\%&8.64\%\\
&Winter&1391&1271&0.13&8.48\%&0.22\%&8.63\%\\
\cline{2-8}
\multirow{2}{*}{P2}&Summer&16795&14171&0.48&16.13\%&-0.51\%&15.62\%\\
&Winter&8567&7010&0.34&17.98\%&0.20\%&18.17\%\\
\cline{2-8}
\multirow{2}{*}{SA}&Summer&2313&1991&0.45&13.88\%&0.04\%&13.92\%\\
&Winter&1215&914	&0.52&24.76\%&0.01\%&24.77\%\\
\cline{2-8}
\multirow{2}{*}{SB}&Summer&32911&27439&0.53&17.72\%&-1.09\%&16.63\%\\
&Winter&15645	&12602&0.37&19.34\%&0.11\%&19.45\%\\
\cline{2-8}
\multirow{2}{*}{SF}&Summer&17679&14224&0.51&18.07\%&1.47\%&19.54\%\\
&Winter&8591&7046&0.28&17.10\%	&0.88\%&17.98\%\\
\cline{2-8}
\hline
\end{tabular}
\caption {Average results based on summer and winter month rates in E-19} \label{tab:restab}
\end{table}

The cumulative load shapes given in Figure~\ref{fig:sampleLoadShapes} and the arrival and departure time histograms given in Figures~\ref{fig:startHour} and~\ref{fig:endHour} support these results. These figures suggest that energy charges increase because a large portion of the EV charging sessions end (i.e.~the charger is unplugged) before the system peak period ends. Thus, when coupled with the higher part-peak demand rates, the optimization converges to a result in which the load is shifted from the EV load peak period (9AM-11AM) to the system peak period (12PM-6PM). \\

\begin{figure}
        \centering
        \begin{subfigure}[a]{0.7\textwidth}
                \includegraphics[width=\textwidth]{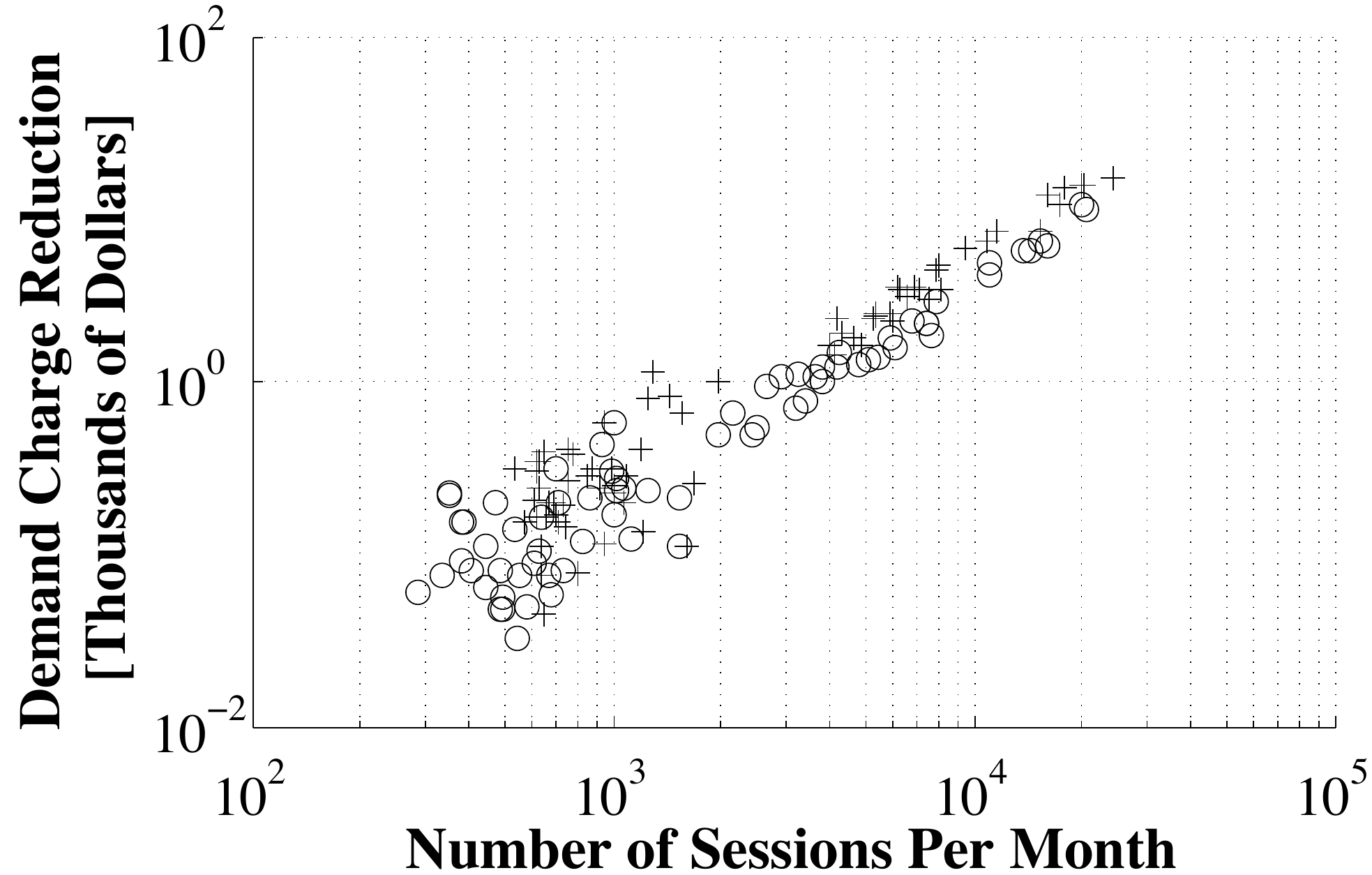}
                \caption{Demand Charge Reduction in Dollars by Session Size}
                \label{fig:sessionDist}
        \end{subfigure}%
        \quad
        \begin{subfigure}[a]{0.7\textwidth}
                \includegraphics[width=\textwidth]{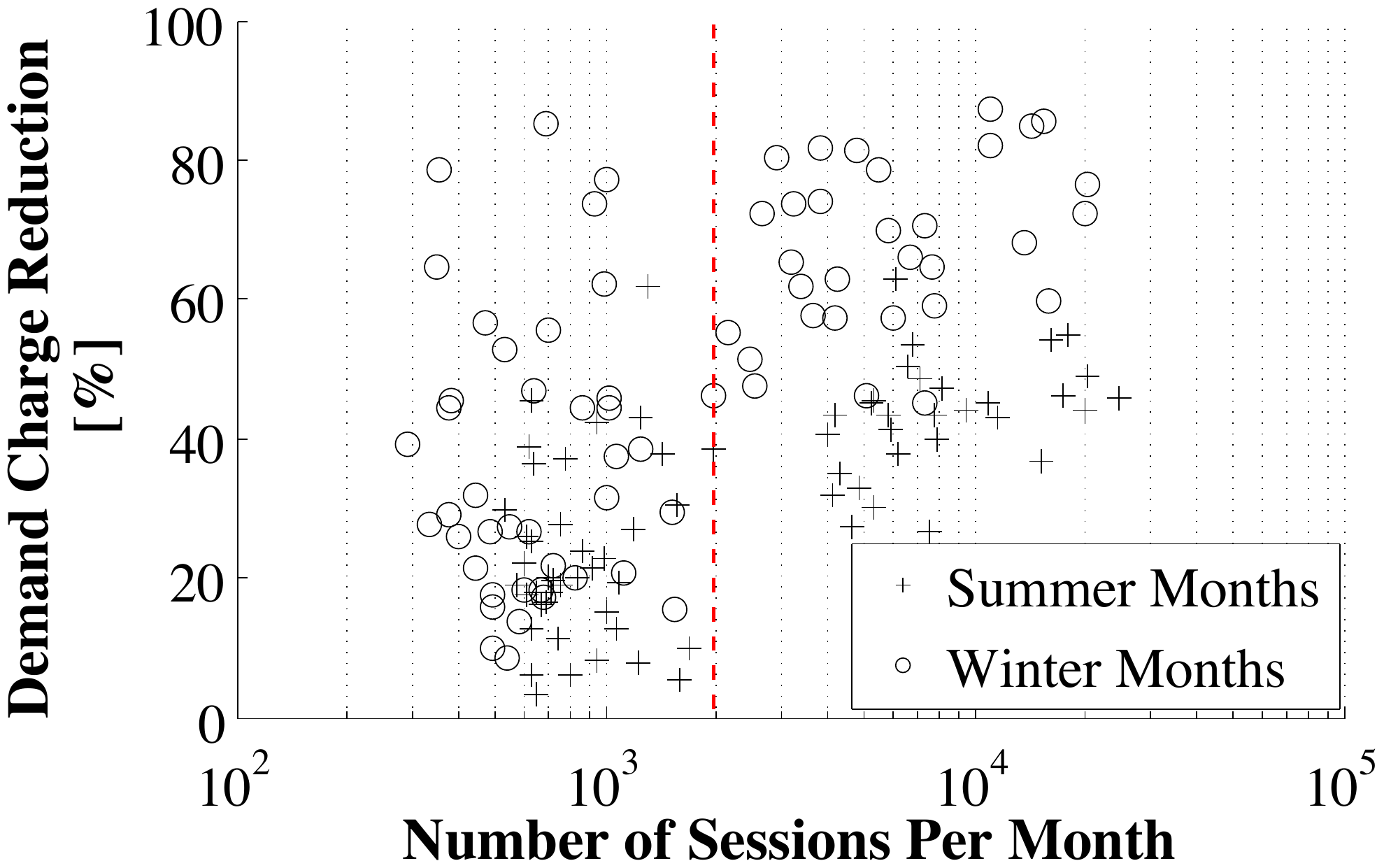}
                \caption{Demand Charge Reduction Percentage by Session Size}
                \label{fig:perDist}
        \end{subfigure}
        \caption{Demand Charge Reduction by Session Size}
\end{figure}

The results given in Table~\ref{tab:restab} provide further insight into the results depicted in Figures~\ref{fig:MonthlyBills}, \ref{fig:MonthlyEnergyCharges} and~\ref{fig:MonthlyDemandCharges}. Specifically, we reflect on the average monthly bill before and after optimization for winter and summer months. Then, we report on average bill reduction per session during these periods. The values range between 0.13 and 0.65 dollars among all VAPs. Overall, we find that the rate structure in the summer periods yields to more reductions per session than the rates in winter months, with the exception of the Sacramento Valley (SA) VAP. We also report on the total percent bill reduction and we break down this percentage into contributions from demand charges and energy charges. We observe that average percent bill reductions range between 8.63\% and 24.77\%. Even though the average reduction per session values are mostly higher during summer months, the relative cost reduction in monthly bills for individual VAPs varies less. This is due to high overall costs in the summer months. \\

Figures~\ref{fig:sessionDist} and~\ref{fig:perDist} depict the relationship between the reduction in demand charges and the number of charging sessions in each VAP per month. Specifically, in Figure~\ref{fig:sessionDist}, we examine the decrease in demand charges in dollars. We observe a linear trend: as the number of sessions per month rises, the reduction in demand charges increases linearly. Given the current load flexibility and arrival and departure times, this is expected because most of the EVs contribute to the peak load of the EV aggregation. In Figure~\ref{fig:perDist}, we look at the percent reduction in demand charges. For up to 2000 charging sessions per month (indicated by a red dashed line in Figure~\ref{fig:perDist}), there is no clear separation between the winter and summer months and, for a given number of sessions, the demand charge reduction values vary. Beyond this point, we can see a clear separation between the winter and summer months, and the demand charge reduction values show less variance.\\

The relative decrease in the summer months is less than the relative decrease in the winter months. We believe that this is due to the time of the peak EV load, the arrival and departure patterns of the EVs and the corresponding rate structure. In particular, the peak EV load coincides with the part-peak rate period, and most of the EVs depart before the system peak period (12PM-6PM) is over. The system peak period has a separate and higher demand rate in the summer months (detailed in Table~\ref{tab:rates}). This limits the smart charging framework's ability to move the EV loads from part-peak period to system peak period. The winter rates we use in this study do not include a separate demand rate for the system peak period; rather, the part-peak period extends from 8:30AM-09:30PM. This makes it possible to manage the EV peak load in a more effective way.\\

\section{Distribution System Operator's Perspective} 
\label{evsec:dso}
In the second case study, we evaluate the potential benefits that smart charging can offer to distribution system operators (DSOs). The motivation behind this case study is to investigate the potential of each charging session to decrease its contribution to the peak system demand via smart charging. We first quantify the percentage of peak load shed during the system peak load period (12AM-6PM). We then quantify the amount of energy that is shifted outside the peak period by the EV load aggregation for each month of 2013. Finally, we report on the amount of energy that can be expected to be moved outside of the system peak period per charging session.\\
\subsection{Problem Formulation}
To realize peak shaving, we propose to develop a two-stage optimization. In the first stage, we minimize a bound on the aggregate power consumed by the EVSEs within a VAP during the pre-defined peak period (12AM-6PM) only. We simplify refer to the pre-defined peak period as $pp$, and to simplify the notation introduced earlier, we refer to the aggregate power vector within the peak period as $AP^{(d)}_{pp}$. To implement the initial stage optimization, we define $AP^{(d)}_{bound,pp}$ as a decision variable to represent the proposed bound on the $AP^{(d)}_{pp}$. Then, in the second stage, using the optimal bound as a constraint, we minimize the total energy consumed within the peak period. This implicitly ensures that the energy bill for the customer is decreased or unchanged based on a typical TOU tariff. 
The first part of the optimization can be written as:

\begin{equation*}
\begin{aligned}
& \underset{\bm{X}^{(i)}\negthickspace,~AP_{bound,pp}^{(d)}}{\text{minimize}}
&  AP^{(d)}_{bound,pp}\\
\end{aligned}
\end{equation*}

\noindent subject to~(\ref{eq:constraints}) and the following additional constraints:

\begin{equation}
\label{eq:optim}
\begin{aligned}
& AP^{(d)}_{pp} \le AP^{(d)}_{bound,pp}\\ \end{aligned} 
\end{equation}

\noindent Then, using the optimal $AP^{(d)}_{bound,pp}$ values obtained in the first stage ${\overset{*}{AP}}{}^{(d)}_{bound,pp}$, we can form the second stage as follows:

\begin{equation*}
\begin{aligned}
& \underset{\bm{X}^{(i)}}{\text{minimize}}
& \underset{\forall k\subseteq pp}{\sum} AP^{(d)}_{k}\\
\end{aligned}
\end{equation*}

\noindent subject to~(\ref{eq:constraints}) and the following additional constraints:

\begin{equation}
\label{eq:optim}
\begin{aligned}
& AP^{(d)}_{pp} \le {\overset{*}{AP}}{}^{(d)}_{bound,pp} \\ \end{aligned} 
\end{equation}

\subsection{Case Study}
The motivation behind our second case study is to evaluate the potential of EV aggregations to decrease their contribution to the system peak load via smart charging. As the arrival and departure time histograms given in Figures~\ref{fig:startHour} and~\ref{fig:endHour} suggest, the amount of energy that can be moved outside of the peak period is expected to be low, mostly because most non-residential EV sessions end before the system peak period is over. However, there is potential in using smart charging and exploiting the inherent flexibility in each charging session to decrease the contribution of EVs to the system peak load.

To demonstrate and quantify this potential, we calculated optimal schedules for each VAP-level aggregation using the optimization strategy described in the above section, and obtained percentage of peak shed values and the total energy moved outside of the peak period for every day in each month of 2013. 
\subsection{Results}
\begin{figure}
\centering
\includegraphics[width=0.9\textwidth]{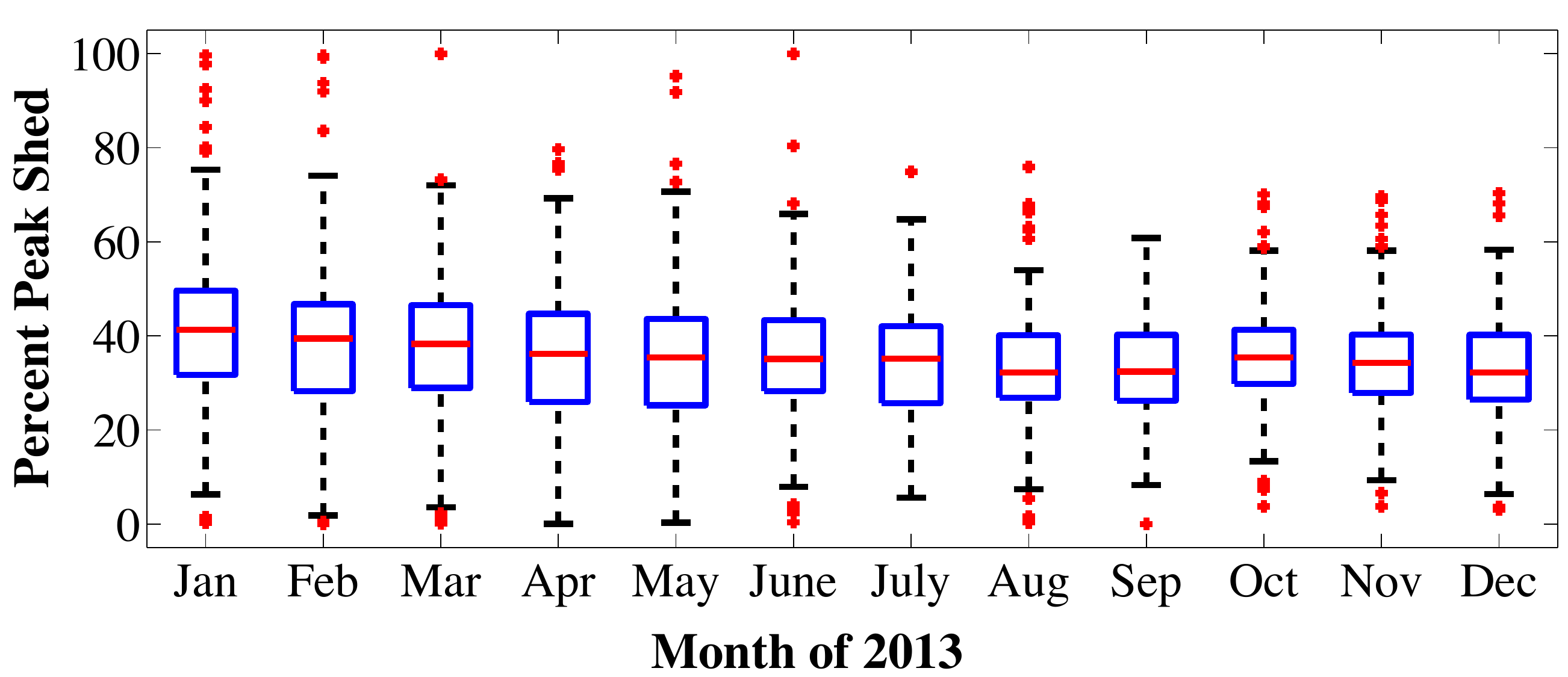}
\caption{Distribution of percent peak shed for all the VAPs}\label{fig:PercentPeakShed}
\end{figure}

Figure~\ref{fig:PercentPeakShed} shows the box plots created using daily peak shed values for each month of 2013. The percentage of peak shed for each day $d$ is defined as: 

\begin{equation}
\%peakshed^{(d)}= \frac{{\stackrel{*}{AP}}{}^{(d)}_{bound,pp}}{max(AP^{(d)}_{pp})}
\end{equation}

The red lines denote the median value of the distribution, the box boundaries are the 25th and 75th percentiles and the whiskers denote the 1st and 99th percentiles, assuming the distributions per each month are normal. The outliers outside the whiskers' boundaries are marked with points. As expected, the smart charging significantly reduces the peak EV load during the system peak period. The median values for all of the months range between 30 and 42\%. A decrease in the peak shaving potential and a slight decrease in the variation of the distributions over the course of 12 months are also apparent in Figure~\ref{fig:PercentPeakShed}. This can be explained by the increase in the number of charging sessions per EVSE and the related decrease in the variation of available flexibility.   \\

\begin{figure}
\centering
\includegraphics[width=0.9\textwidth]{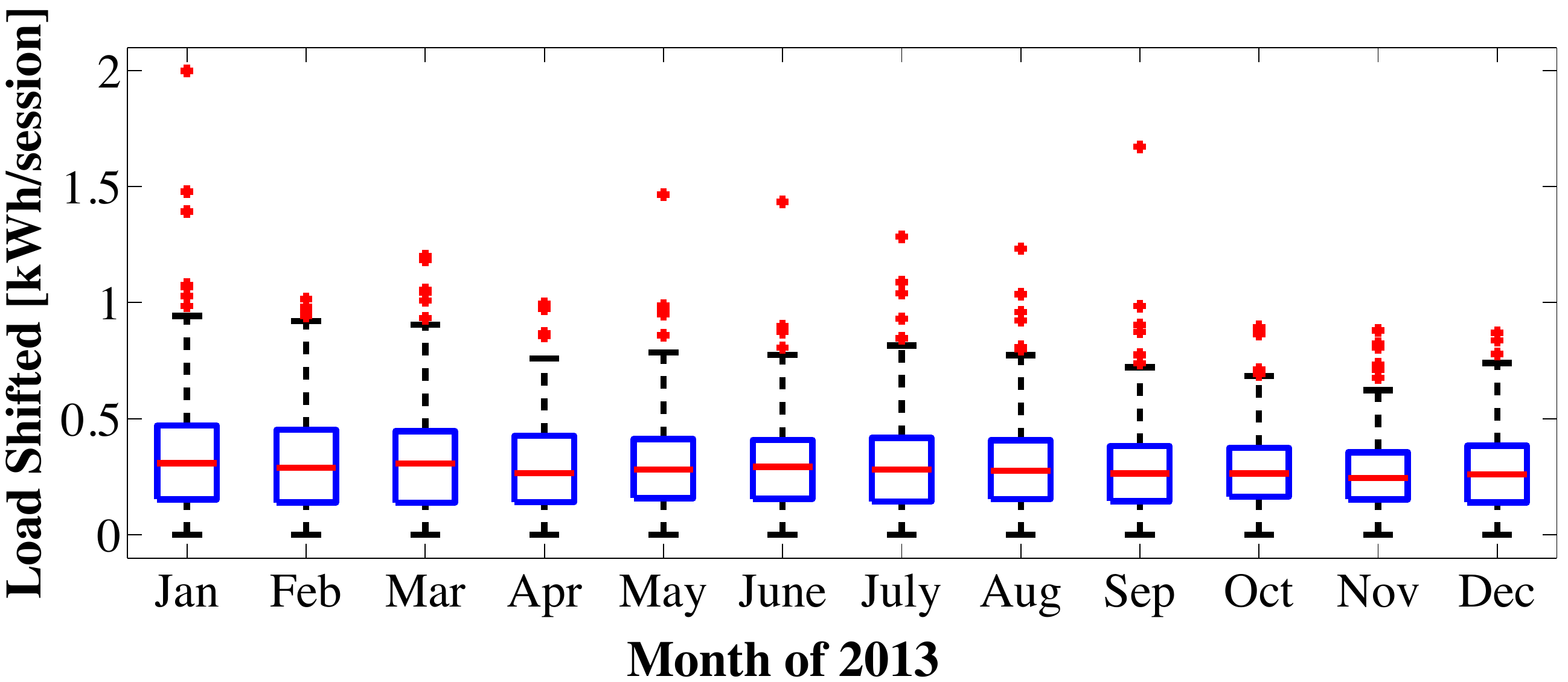}
\caption{Total energy moved outside of system peak period}\label{fig:LoadShiftedOutsidePerSession}
\end{figure}

Figure~\ref{fig:LoadShiftedOutsidePerSession} depicts the distribution of the average energy moved outside of the peak period per charging session for all of the VAPs estimated every day of the month.  
The median value over 2013 is approximately 0.25kWh per charging session, which corresponds to $\sim$2.8\% of the average energy put during each charging session.

\section{Conclusions and Future Work}
\label{evsec:conc}
In this paper, we quantify the potential benefits of smart charging to different stakeholders using data collected from over 2000 non-residential electric vehicle supply equipment (EVSEs) located throughout 190 zip code regions in Northern California. We created virtual aggregation points (VAP) in which the aggregate power consumption of a selected population of EVSEs is assumed to be managed via individual charging control at each EVSE. We developed and used a smart charging framework to estimate the benefits of EV smart charging to different stakeholders:  a single owner/an aggregator of behind-the-meter EVSEs (i.e. aggregators) and distribution system operators.\\

In our first case study, we investigated the potential benefits of behind-the-meter EV aggregations. The aggregate load is re-scheduled using a TOU rate structure. Our results suggest that up to 24.8\% decrease in the aggregate monthly bill per VAP is possible. In all months, this reduction is due to a corresponding decrease in demand charges in the monthly bill: we observed that decreases in energy charges are contributing by up to 1.5\% to the overall decrease, whereas the demand charges contribute up to 24.7\%.\\

In our second case study, we used the EV aggregations to decrease their contribution to the system-level peak load. We have observed median peak shed values around 30\%-42\% for each month. In addition, we have quantified the amount of energy that can be shifted outside the peak period per charging session over the course of 2013, and found the median value to be approximately 0.25kWh/session ($\sim$2.8\% of the average energy put in each session). \\

In the future, we would like to investigate the impact of different non-residential customer categories (e.g., retail vs. workplace) within each VAP to similar metrics calculated in this study and identify suitable grid services for these customer categories. In addition, we would like to expand the current smart charging framework and develop control algorithms for workplace charging that use variable charging rates. We also would like to study the impacts of smart non-residential EV charging to the overall system load, in particular when the system level solar generation is expected to cause over-generation and ramping problems in the grid. 

\section*{Acknowledgments} 
\addcontentsline{toc}{section}{Acknowledgment} 
We would like to thank Pacific Gas and Electric Company and ChargePoint LLP for providing the data used in this study. We would also like to thank Salman Masyakeh for helpful discussions.  This research was supported in part by the Pennsylvania Infrastructure Technology Alliance.
\section*{References}
\bibliographystyle{elsarticle-num}
\bibliography{sgc}

\begin{thebibliography}{10}
\expandafter\ifx\csname url\endcsname\relax
  \def\url#1{\texttt{#1}}\fi
\expandafter\ifx\csname urlprefix\endcsname\relax\def\urlprefix{URL }\fi
\expandafter\ifx\csname href\endcsname\relax
  \def\href#1#2{#2} \def\path#1{#1}\fi

\bibitem{williams2012technology}
J.~H. Williams, A.~DeBenedictis, R.~Ghanadan, A.~Mahone, J.~Moore, W.~R.
  Morrow, S.~Price, M.~S. Torn, The technology path to deep greenhouse gas
  emissions cuts by 2050: the pivotal role of electricity, science 335~(6064)
  (2012) 53--59.

\bibitem{duvall_environmental_2007}
M.~Duvall, E.~Knipping, M.~Alexander, L.~Tonachel, C.~Clark, Environmental
  assessment of plug-in hybrid electric vehicles, EPRI, July.

\bibitem{liu_survey_2011}
R.~Liu, L.~Dow, E.~Liu, A survey of {PEV} impacts on electric utilities, in:
  Innovative Smart Grid Technologies (ISGT), 2011 IEEE PES, IEEE, 2011, pp.
  1--8.

\bibitem{wu2011electric}
D.~Wu, D.~Aliprantis, K.~Gkritza, Electric energy and power consumption by
  light-duty plug-in electric vehicles, Power Systems, IEEE Transactions on
  26~(2) (2011) 738--746.

\bibitem{weiller_plug_2011}
C.~Weiller,
  \href{http://www.sciencedirect.com/science/article/pii/S0301421511002886}{Plug-in
  hybrid electric vehicle impacts on hourly electricity demand in the united
  states}, Energy Policy 39~(6) (2011) 3766 -- 3778.
\newblock \href
  {http://dx.doi.org/http://dx.doi.org/10.1016/j.enpol.2011.04.005}
  {\path{doi:http://dx.doi.org/10.1016/j.enpol.2011.04.005}}.
\newline\urlprefix\url{http://www.sciencedirect.com/science/article/pii/S0301421511002886}

\bibitem{lopes_integration_2011}
J.~Lopes, F.~Soares, P.~Almeida, Integration of electric vehicles in the
  electric power system, Proceedings of the IEEE 99~(1) (2011) 168--183.
\newblock \href {http://dx.doi.org/10.1109/JPROC.2010.2066250}
  {\path{doi:10.1109/JPROC.2010.2066250}}.

\bibitem{galus_role_2012}
M.~D. Galus, M.~G. Vayá, T.~Krause, G.~Andersson,
  \href{http://dx.doi.org/10.1002/wene.56}{The role of electric vehicles in
  smart grids}, Wiley Interdisciplinary Reviews: Energy and Environment 2~(4)
  (2013) 384--400.
\newblock \href {http://dx.doi.org/10.1002/wene.56}
  {\path{doi:10.1002/wene.56}}.
\newline\urlprefix\url{http://dx.doi.org/10.1002/wene.56}

\bibitem{guille2009conceptual}
C.~Guille, G.~Gross, A conceptual framework for the vehicle-to-grid
  implementation, Energy Policy 37~(11) (2009) 4379--4390.

\bibitem{kempton2005vehicle}
W.~Kempton, J.~Tomi{\'c}, Vehicle-to-grid power implementation: From
  stabilizing the grid to supporting large-scale renewable energy, Journal of
  Power Sources 144~(1) (2005) 280--294.

\bibitem{brooks2001integration}
A.~Brooks, T.~Gage, A.~Propulsion, Integration of electric drive vehicles with
  the electric power grid—a new value stream, in: 18th International Electric
  Vehicle Symposium and Exhibition, Berlin, Germany, Citeseer, 2001, pp.
  20--24.

\bibitem{martin2012direct}
P.~Sanchez-Martin, G.~Sanchez, G.~Morales-Espana, Direct load control decision
  model for aggregated ev charging points, Power Systems, IEEE Transactions on
  27~(3) (2012) 1577--1584.
\newblock \href {http://dx.doi.org/10.1109/TPWRS.2011.2180546}
  {\path{doi:10.1109/TPWRS.2011.2180546}}.

\bibitem{su2012performance}
W.~Su, M.-Y. Chow, Performance evaluation of an eda-based large-scale plug-in
  hybrid electric vehicle charging algorithm, Smart Grid, IEEE Transactions on
  3~(1) (2012) 308--315.

\bibitem{he2012optimal}
Y.~He, B.~Venkatesh, L.~Guan, Optimal scheduling for charging and discharging
  of electric vehicles, Smart Grid, IEEE Transactions on 3~(3) (2012)
  1095--1105.

\bibitem{rotering_2011_optimal}
N.~Rotering, M.~Ilic, Optimal charge control of plug-in hybrid electric
  vehicles in deregulated electricity markets, Power Systems, IEEE Transactions
  on 26~(3) (2011) 1021--1029.
\newblock \href {http://dx.doi.org/10.1109/TPWRS.2010.2086083}
  {\path{doi:10.1109/TPWRS.2010.2086083}}.

\bibitem{tomic2007using}
J.~Tomi{\'c}, W.~Kempton, Using fleets of electric-drive vehicles for grid
  support, Journal of Power Sources 168~(2) (2007) 459--468.

\bibitem{kempton2008test}
W.~Kempton, V.~Udo, K.~Huber, K.~Komara, S.~Letendre, S.~Baker, D.~Brunner,
  N.~Pearre, A test of vehicle-to-grid (v2g) for energy storage and frequency
  regulation in the pjm system, Results from an Industry-University Research
  Partnership (2008) 32.

\bibitem{finn2012demand}
P.~Finn, C.~Fitzpatrick, D.~Connolly,
  \href{http://www.sciencedirect.com/science/article/pii/S0360544212002435}{Demand
  side management of electric car charging: Benefits for consumer and grid},
  Energy 42~(1) (2012) 358 -- 363, 8th World Energy System Conference, \{WESC\}
  2010.
\newblock \href
  {http://dx.doi.org/http://dx.doi.org/10.1016/j.energy.2012.03.042}
  {\path{doi:http://dx.doi.org/10.1016/j.energy.2012.03.042}}.
\newline\urlprefix\url{http://www.sciencedirect.com/science/article/pii/S0360544212002435}

\bibitem{sundstrom_flexible_2012}
O.~Sundstrom, C.~Binding, Flexible charging optimization for electric vehicles
  considering distribution grid constraints, Smart Grid, IEEE Transactions on
  3~(1) (2012) 26--37.

\bibitem{dietz_economic_2011}
B.~Dietz, K.~Ahlert, A.~Schuller, C.~Weinhardt, Economic benchmark of charging
  strategies for battery electric vehicles, in: PowerTech, 2011 IEEE Trondheim,
  IEEE, 2011, pp. 1--8.

\bibitem{deilami_realtime_2011}
S.~Deilami, A.~S. Masoum, P.~S. Moses, M.~A. Masoum, Real-time coordination of
  plug-in electric vehicle charging in smart grids to minimize power losses and
  improve voltage profile, Smart Grid, IEEE Transactions on 2~(3) (2011)
  456--467.

\bibitem{shaaban_real_2014}
M.~Shaaban, M.~Ismail, E.~El-Saadany, W.~Zhuang, Real-time pev
  charging/discharging coordination in smart distribution systems, Smart Grid,
  IEEE Transactions on 5~(4) (2014) 1797--1807.
\newblock \href {http://dx.doi.org/10.1109/TSG.2014.2311457}
  {\path{doi:10.1109/TSG.2014.2311457}}.

\bibitem{marra_demand_2012}
F.~Marra, G.~Y. Yang, C.~Traholt, E.~Larsen, C.~N. Rasmussen, S.~You, Demand
  profile study of battery electric vehicle under different charging options,
  in: Power and Energy Society General Meeting, 2012 IEEE, IEEE, 2012, pp.
  1--7.

\bibitem{optimization2014inc}
G.~Optimization, Inc.: Gurobi optimizer reference manual (2012) (2014).

\bibitem{e19}
{Pacific Gas and Electric Company},
  \href{http://www.pge.com/tariffs/tm2/pdf/ELEC_SCHEDS_E-19.pdf}{{Electric
  Schedule E-19:Medium General Demand-Metered {TOU} Service}} (2010).
\newline\urlprefix\url{http://www.pge.com/tariffs/tm2/pdf/ELEC_SCHEDS_E-19.pdf}

\end{thebibliography}

\end{document}